\documentclass[a4,12pt,draftclsnofoot,onecolumn]{IEEEtran}

\usepackage[T1]{fontenc}

\usepackage[none]{hyphenat}
\usepackage[utf8]{inputenc}
\usepackage[english]{babel}

\usepackage{verbatim}
\usepackage{graphicx}
\usepackage{subfigure}
\usepackage{float}
\usepackage{hyperref}
\usepackage{lipsum}
\usepackage{stfloats}
\usepackage{amsmath,amsthm}
\usepackage{amssymb}
\usepackage{mathtools}
\usepackage{epstopdf}
\usepackage{booktabs,dcolumn,caption}
\usepackage{multirow}
\usepackage{caption}
\usepackage{ragged2e}
\usepackage{cite}
\usepackage{censor}

\usepackage{array}
\newcolumntype{C}[1]{>{\centering\arraybackslash}p{#1}}
\newcolumntype{L}[1]{>{\RaggedRight\arraybackslash}p{#1}}
\usepackage{wasysym}

\newcommand*\diff{\mathop{}\!\mathrm{d}}

\usepackage{mathrsfs}
\usepackage{xcolor}
\usepackage{colortbl}
\usepackage{color,soul}

\usepackage{caption,hypcap}

\usepackage{cleveref}

\usepackage{suffix}

\usepackage{cleveref}
\usepackage{scalerel}
\usepackage{tikz}
\usetikzlibrary{svg.path}

\definecolor{orcidlogocol}{HTML}{A6CE39}
\tikzset{
  orcidlogo/.pic={
    \fill[orcidlogocol] svg{M256,128c0,70.7-57.3,128-128,128C57.3,256,0,198.7,0,128C0,57.3,57.3,0,128,0C198.7,0,256,57.3,256,128z};
    \fill[white] svg{M86.3,186.2H70.9V79.1h15.4v48.4V186.2z}
                 svg{M108.9,79.1h41.6c39.6,0,57,28.3,57,53.6c0,27.5-21.5,53.6-56.8,53.6h-41.8V79.1z M124.3,172.4h24.5c34.9,0,42.9-26.5,42.9-39.7c0-21.5-13.7-39.7-43.7-39.7h-23.7V172.4z}
                 svg{M88.7,56.8c0,5.5-4.5,10.1-10.1,10.1c-5.6,0-10.1-4.6-10.1-10.1c0-5.6,4.5-10.1,10.1-10.1C84.2,46.7,88.7,51.3,88.7,56.8z};
  }
}

\newcommand\orcidicon[1]{\href{https://orcid.org/#1}{\mbox{\scalerel*{
\begin{tikzpicture}[yscale=-1,transform shape]
\pic{orcidlogo};
\end{tikzpicture}
}{|}}}}

\usepackage{mathtools, cuted}
\usepackage{lipsum, color}

\DeclarePairedDelimiterX\MeijerM[3]{\lparen}{\rparen}%
{#3\delimsize\vert\,\begin{matrix}#1 \\ #2\end{matrix}}

\newcommand\MeijerG[8][]{%
	G^{\,#2,#3}_{#4,#5}\MeijerM[#1]{#6}{#7}{#8}}

\WithSuffix\newcommand\MeijerG*[7]{%
	G^{\,#1,#2}_{#3,#4}\MeijerM*{#5}{#6}{#7}}

\begin{document}

\title{Moment-based Parameter Estimation for \texorpdfstring{$\Gamma$}{Γ}-parameterized TWDP Model}

\author{
        Pamela~Njemcevic
        ,~\IEEEmembership{Member,~IEEE,}
        Almir~Maric
        ,~\IEEEmembership{Member,~IEEE,}
        and~Enio~Kaljic
        ,~\IEEEmembership{Member,~IEEE}
\thanks{P. Njemcevic, A. Maric and E. Kaljic are with the Department
of Telecommunications, Faculty of Electrical Engineering, University of Sarajevo, Sarajevo, Bosnia and Herzegovina.}
\thanks{e-mail: pamela.njemcevic@etf.unsa.ba;}
}

\maketitle

\begin{abstract}
In this paper, parameterization of a two-wave with diffuse power (TWDP) fading model is revised. 
Anomalies caused by using conventional TWDP parameters $K$ and $\Delta$ are first identified, indicating that
the existing moment-based estimators of a tuple $(K, \Delta)$ are not able to provide accurate estimations for various combinations of their values.
Therefore, a moment-based estimators for a newly proposed parameters $K$ and $\Gamma$ are derived and analyzed through asymptotic variance (AsV) and Cramer-Rao bound (CRB) metrics. The results are qualitatively compared to those obtained for a $(K, \Delta)$ tuple, showing that moment-based estimators of the improved $\Gamma$-based parameterization managed to overcome all anomalies observed within the  conventional TWDP parameterization. 
\end{abstract}

\section{Introduction}
\label{Sec:1}
Over the last two decades, TWDP model has been extensively used to characterize  small-scale fluctuations of a signal envelope, in cases where mmWave band and directional antennas are employed~\cite{Zoc19, Zoc19-1, Mav15}, as well as in wireless sensor networks within the cavity environments~\cite{Fro07}. In these conditions, the waves arriving at the receiver can be observed as a sum of two specular line-of-site (LOS) components with constant magnitudes $V_1$ and $V_2$ and uniformly distributed phases, plus many diffuse non-LOS components treated as a complex zero-mean Gaussian random process with average power $2\sigma^2$. As such, the model is conventionally characterized by the parameters $K$, $\Delta$, and $\Omega$ defined as~\cite{Dur02}:  
\begin{equation}
\label{eq1}
   K = \frac{V_1^2+V_2^2}{2 \sigma^2}, ~~ \Delta = \frac{2V_1 V_2}{V_1^2+V_2^2}, ~~ \Omega = V_1^2 + V_2^2 + 2\sigma^2
\end{equation}
where parameter $K$ ($K \geq 0$) characterizes the ratio of the average power of the specular components to the power of the remaining diffuse components (like the Rician parameter $K$), parameter $\Delta$ \mbox{$(0\leq \Delta \leq 1)$}  characterizes the relation between magnitudes of specular components, and $\Omega$ represents the average power of the received signal.

However, it is elaborated in~\cite{rad} that definition of  parameter $\Delta$ is not in accordance with model’s underlying physical mechanisms. 
Namely, according to the model's definition, specular components have constant magnitudes and are propagating in a linear medium. So, the function which characterizes the 
ratio between $V_1$ and $V_2$ has to be linear~\cite{rad}. On the other hand, parameter $\Delta$ introduces nonlinear relation between the magnitude of specular components (since \mbox{$V_2 = V_1(1-\sqrt{(1-\Delta^2)})/\Delta$}), hindering accurate observation of the impact of their ratio on a system's performance metrics~\cite{rad}. 
Therefore, a novel parameterization proposed in~\cite{rad} introduces parameter $\Gamma$ instead of $\Delta$, defined as: 
\begin{equation}
\label{eq2}
\Gamma = \frac{V_2}{V_1}
\end{equation}
where $0 \leq \Gamma \leq 1$ for $0 \leq V_2 \leq V_1$ obviously ensures linear dependence between $V_1$ and $V_2$. According to~\cite{rad}, the  definition of parameter $K$ in novel TWDP parameterization remains unchanged. However, it can be observed that parameter $\Delta$ completely changes the character of the original definition expression of parameter $K$ in (\ref{eq1}), since $K$ vs. $\Delta$ can be expressed as $K = (V_1^2+V_2^2)/(2\sigma^2)=({V_1^2}/{2\sigma^2})({2}/{\Delta})({V_2}/{V_1})= 2 K_{Rice}/({1-\sqrt{1-\Delta^2}}){\Delta^2}$
\noindent where $K_{Rice} = V_1/2\sigma^2$. On the other side, when $K$ is expressed in terms of $\Gamma$, as $K = {V_1^2}/{2 \sigma^2}\left[1+({V_2}/{V_1})^2\right] = K_{Rice}(1 + \Gamma^2)$, 
the character of the original definition expression of $K$ in (\ref{eq1}) given in terms of $V_2/V_1$, remains unchanged by the parameter $\Gamma$. 

Beside the aforesaid, nonphysical definition of parameter $\Delta$ also causes anomalies related to the estimation of $K$ and $\Delta$. These anomalies are for the first time observed in~\cite{Fer16}, by noticing "\textit{when $\Delta$
is an unknown parameter, the estimation error bound is lower for higher $\Delta$ values ($\Delta \to 1$). This is in contrast to the opposite situation on which $\Delta$ is a known parameter, and parameter $K$ is estimated}". 
However, observed anomaly has not been further investigated. Another anomaly can be observed in~\cite[Fig. 4]{Fer16-1}, where estimated values of parameter $\Delta$ (i.e. $\hat{\Delta}$), obtained from $500$ different realizations of TWDP process with $N = 10^4$ i.i.d. samples, are illustrated along with the sample means of these estimated values for each considered tuple $(K, \Delta)$. The figure clearly shows that $\hat{\Delta}$ 
takes values greater than one for $\Delta \approx 1$.
However, according to definition given by (\ref{eq1}), parameter $\Delta$ can take values only between zero and one, indicating that the estimates greater than one are nonphysical and useless for gaining insight into the relation between $V_1$ and $V_2$.
The consequence of the above can be observed in sample mean of estimated values in the vicinity of one, for which obviously fake accurate results are obtained by averaging potentially accurate values of $\hat{\Delta}$ ($\hat{\Delta} \leq 1$) and incorrect ones ($\hat{\Delta} > 1$). 
That caused decrements in estimation error of $\Delta$ as it approaches one (as shown in~\cite[Fig. 2]{Fer16-1}) and occurrence of the anomaly recognized in~\cite{Fer16} as inadequate behavior of $\hat{\Delta}$. 
Therefore, although~\cite{Fer16, Fer16-1} provide mathematically correct 
expressions for estimators of $K$ and $\Delta$, the results obtained by their application are inaccurate due to nonphysical definition of $\Delta$. 



So, after the anomalies caused by conventional parameterization are identified, an overview of TWDP parameters' estimation results is presented, indicating the absence of those related to the estimation of a tuple $(K, \Gamma)$. Thus, a closed-form moment-based estimators of parameters $K$ and $\Gamma$ are derived and
analyzed through their AsVs and  CRBs. Estimators of conventional and improved parameters are then compared, but only in order to gain qualitative insight into the differences in their behaviour. Otherwise, due to described problems in definition of $\Delta$ and its impact on the definition of $K$, it makes no sense to perform quantitative comparison between the results obtained for the tuples $(K, \Gamma)$ and $(K, \Delta)$. 

\section{TWDP parameters' estimation - an overview}
\label{Sec:2}
The estimation of TWDP parameters is of a practical importance in a variety of wireless scenarios. It includes not only delay insensitive channel characterization and link budget calculations, but also online adaptive coding/modulation for which estimation of parameters must be both accurate and prompt. Accordingly, different approaches in estimation of TWDP parameters are proposed to make a deal between the computational complexity and estimation accuracy. These approaches are used to estimate a tuple $(K, \Delta)$, while the parameter $\Omega$ has not been usually the part of the approaches (since it can be directly estimated from the data set as a second moment)~\cite{Zoc19-1}. 

Among the investigated approaches, the distribution fitting approach is used to estimate a tuple $(K, \Delta)$ from measurements performed in air-crafts and buses at 2.4 GHz~\cite{Fro07}, while the maximum likelihood procedure (ML) is used to estimate the tuple $(K, \Delta)$ at 60 GHz in indoor environment~\cite{Zoc19} and in vehicular-to-vehicular propagation scenario~\cite{Zoc19-1}. However, it is shown that both approaches are very computationally complex and inappropriate for online applications. Accordingly, the moment-based approach is considered in~\cite{Fer16, Fer16-1, Mav15} as a compromise between the complexity and the estimation accuracy. Thereby, in~\cite{Fer16, Fer16-1}, analytical expressions for $K$ and $\Delta$ estimators are derived and examined in terms of AsV and CRB. In~\cite{Fer16}, parameters $K$ and $\Delta$ are estimated separately under the assumption that one of them is previously known. However, although providing valuable insight into behaviour of $(K,\Delta)$ estimators, expressions from~\cite{Fer16} can not be used for empirical estimation, since in reality both parameters are unknown. The issue is overcome in~\cite{Fer16-1} by deriving computationally simple joint estimators of parameters $(K, \Delta)$. 

\begin{figure*}[h!]
\centering
\begin{minipage}[t]{1\textwidth}
  \centering
  \includegraphics[trim={0.6cm 0.2cm 1.2cm 1.8cm},clip,width=0.7\linewidth]{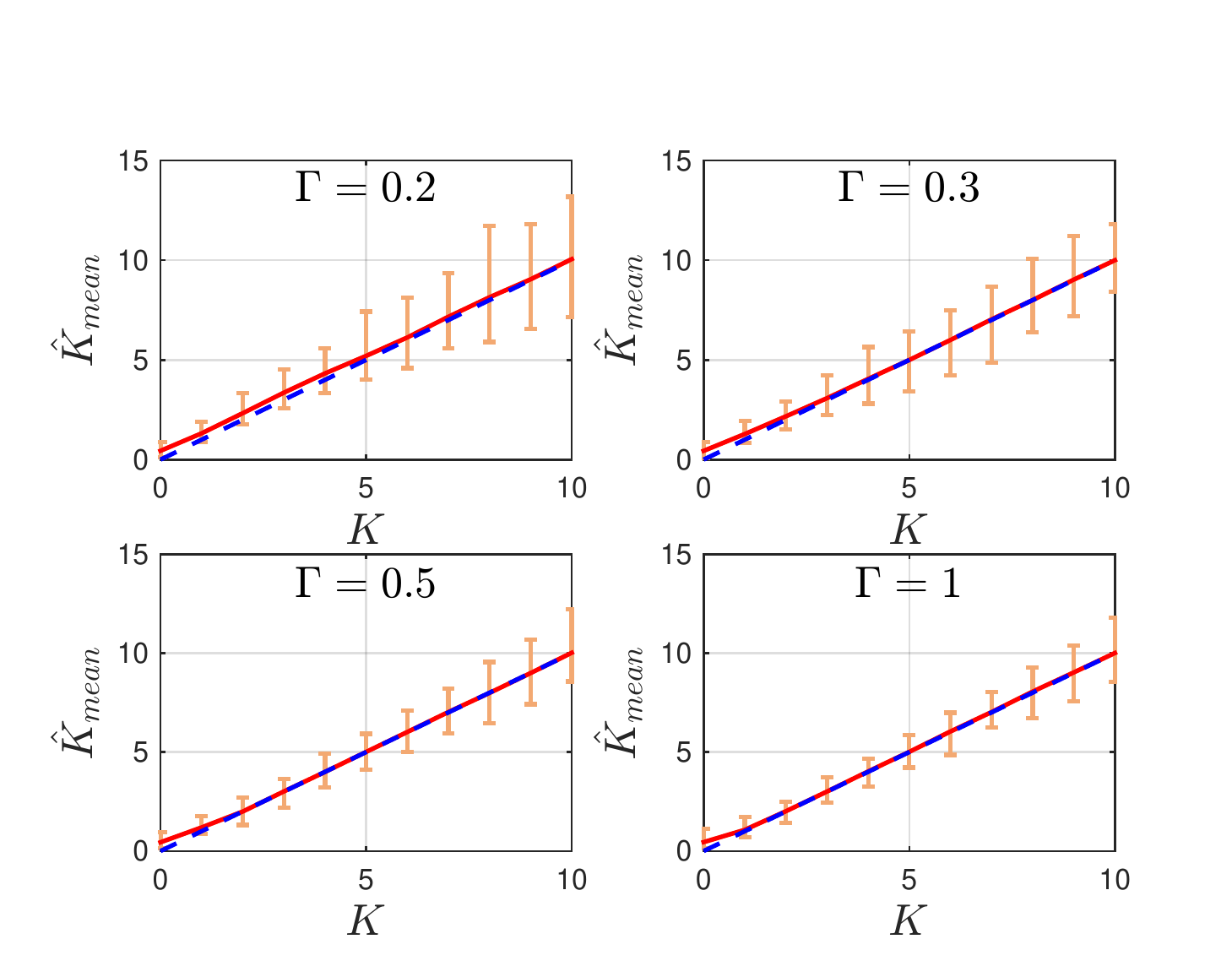}
  \caption{$\hat{K}_{mean}$  (with  absolute  error  bars)  vs. $K$ for different values of $\Gamma$. The solid line shows a linear regression fit to the data. Unit slope dashed line is illustrated as a benchmark.}
  \label{Figure_nova}
\end{minipage}%
\vspace{0.7cm}
\begin{minipage}[t]{1\textwidth}
  \centering
  \includegraphics[trim={0.6cm 0.2cm 1.2cm 1.8cm},clip,width=0.7\linewidth]{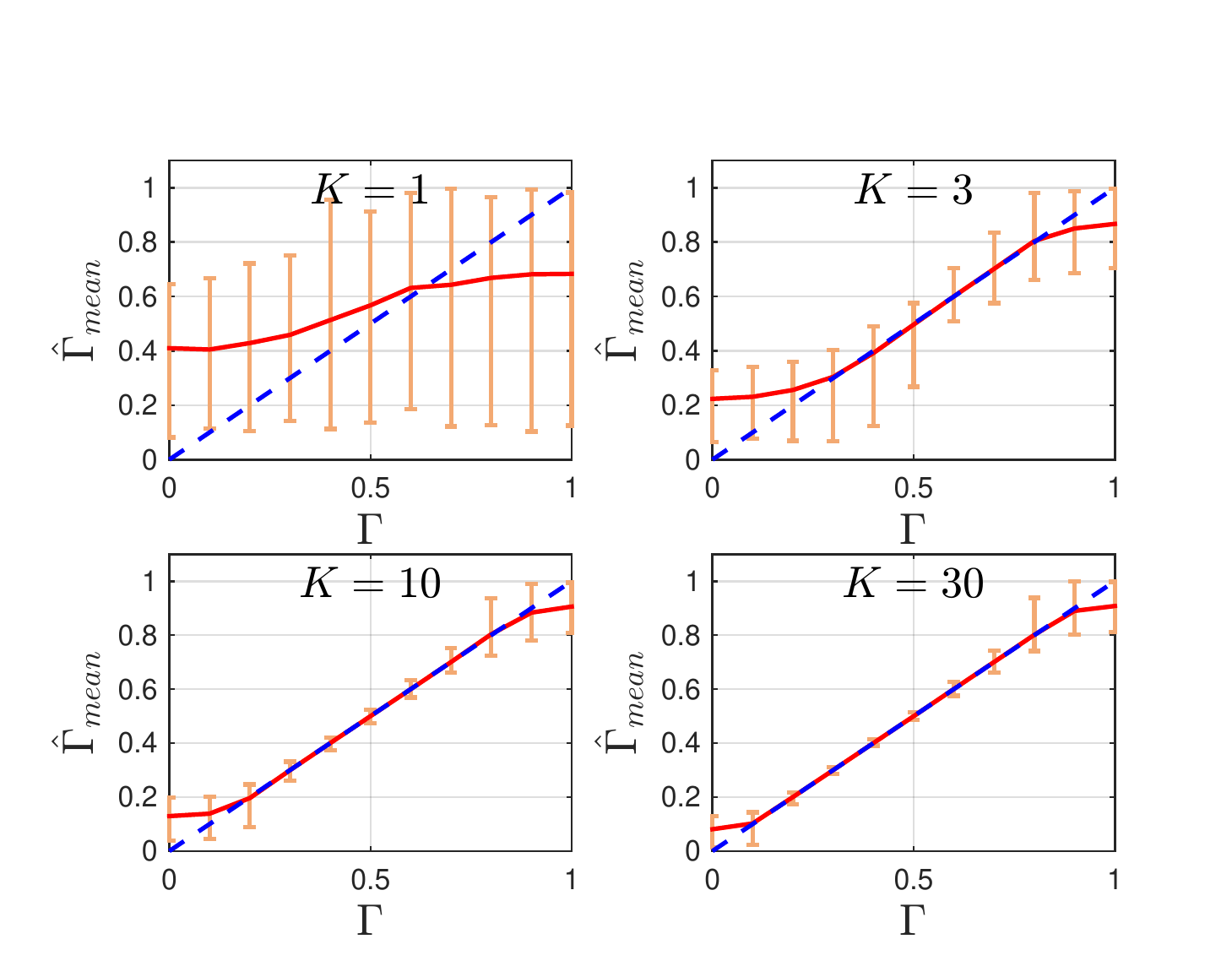}
  \caption{$\hat{\Gamma}_{mean}$  (with  absolute error bars)  vs. $\Gamma$ for different values of $K$. The solid line shows a linear regression fit to the data. Unit slope dashed line is illustrated as a benchmark.}
  \label{Figure_dd}
\end{minipage}%
\end{figure*}

However, due to nonphysical definition of parameter $\Delta$, estimators derived in~\cite{Fer16-1} provide irrelevant results for some combinations of parameters $K$ and $\Delta$. Therefore, it is necessary to derive estimator for physically justified TWDP parameters $K$ and $\Gamma$ and to investigate their behaviour in terms of asymptotic efficiency and the estimation accuracy.

\section{Moment-based estimation of the improved TWDP parameters}
\label{Sec:3}
To create moment-based estimators for improved set of TWDP parameters, the general expression for even moments of the signal envelope $r$ is derived from the expression for moments of a squared signal envelope $\gamma = r^2$~\cite[eq. (6)]{Fer16-1}, by transforming parameter $\Delta$ to $\Gamma$, ($\Delta=2\Gamma/(1+\Gamma^2)$), as:
\begin{equation}
    \label{eq_3}
    \begin{split}
    \mu_n & = \mathbb{E}[r^n] =  \mathbb{E}[{\gamma}^{\frac{n}{2}}] =  \frac{\left({\frac{n}{2}}\right)!{\Omega}^{\frac{n}{2}}}{(1+K)^{\frac{n}{2}} 2\pi} \sum_{m = 0}^{\frac{n}{2}} \binom{{\frac{n}{2}}}{m} \frac{K^m}{m!} \\
    & \times \int_{0}^{2\pi} \left(1 + \left(\frac{2\Gamma}{1 + \Gamma^2} \right) \cos(\theta)\right)^m \diff{\theta}, ~~ \text{for}~~\frac{n}{2} \in \mathbb{N}
\end{split}
\end{equation}
Obviously, the signal envelope's $n$-th moments $\mu_n = \mathbb{E}[r^n]$ depends on three unknown parameters: $K$, $\Gamma$, and $\Omega$. Consequently, the TWDP parameters' estimators can be constructed from at least three different moments. 
Thereby, it is shown that the estimation accuracy is the largest when the lower-order moments are used~\cite{Fer16-1}. However, since only even moments are obtained as closed-form expressions, moment-based estimators of improved TWDP parameters are generated using second-, fourth-, and sixth-order moments. 
To further reduce the estimation complexity, the impact of parameter $\Omega$ on $K$ and $\Gamma$ is  canceled out by properly defining ratios between the fourth- and second-order as well as the sixth- and second-order moments of an envelope~\cite{Fer16-1}, 
as:

\begin{equation}
    \label{eq_4}
    \frac{\mu_4}{\mu_2^2} = \frac{\left(2 + \left(\frac{2\Gamma}{1 + \Gamma^2}\right)^2\right)K^2 + 8K + 4}{2(1+K)^2}
\end{equation}
\begin{equation}
    \begin{split}
    \frac{\mu_6}{\mu_2^3} = \frac{\left(6 + 9\left(\frac{2\Gamma}{1 + \Gamma^2}\right)^2\right)K^3 + \left(42 + 9\left(\frac{2\Gamma}{1 + \Gamma^2}\right)^2\right)K^2 }{2(1+K)^3} + \frac{36K + 12}{2(1+K)^3} 
    \end{split}
    \nonumber
\end{equation}

\vspace{-0.3cm}
\noindent Using the sample moments $\hat{\mu}_n = \frac{1}{N}\sum_{i = 1}^N{r_i^n}$ instead of the ensemble averages $\mu_n$, the system (\ref{eq_4}) is solved for $K$ and $\Gamma$ using results from~\cite{Fer16-1} and the established relation between $\Delta$ and $\Gamma$, 
providing expressions for moment-based estimators $\hat{K}$ and $\hat{\Gamma}$ as:
\begin{equation}
\label{eq_6}
    \hat{K} =  \left(p+\sqrt{p^2+q^3}\right)^\frac{1}{3} +{\left(p-\sqrt{p^2+q^3}\right)}^\frac{1}{3} - \frac{a_1}{3}
\end{equation}
\begin{equation}
    \label{eq_7}
    \hat{\Gamma} = \frac{
            \sqrt{2} \hat{K} \left(
                1 - \sqrt{\left( \frac{2 \left( 4\hat{K} + \hat{K}^2 - \frac{\hat{\mu}_4}{\hat{\mu}_2^2}(\hat{K} + 1)^2 + 2\right)}{\hat{K}^2} \right)+1}
            \right)
        }{
            2\sqrt{\frac{\hat{\mu}_4}{\hat{\mu}_2^2}(\hat{K}+1)^2-\hat{K}^2 - 4\hat{K} - 2}
        }
\end{equation}
where $p$, $q$, $a_1$, $a_2$, and $a_3$ are defined in~\cite{Fer16-1} as:
\[
    p = \frac{1}{54}(9a_1 a_2-27 a_3-2 a_1^3), ~~~ q = \frac{1}{9}(3a_2-a_1^2)
\]
\[
    a_1 = \frac{6\hat{\mu}_6 - 30\hat{\mu}_4\hat{\mu}_2 + 24\hat{\mu}_2^3}{2\hat{\mu}_6 - 6\hat{\mu}_4\hat{\mu}_2 + 4\hat{\mu}_2^3}  ~~~~~~~~~  a_2 = \frac{6\hat{\mu}_6 - 42\hat{\mu}_4\hat{\mu}_2 + 48\hat{\mu}_2^3}{2\hat{\mu}_6 - 6\hat{\mu}_4\hat{\mu}_2 + 4\hat{\mu}_2^3}
\]
\[
    a_3 = \frac{2\hat{\mu}_6 - 18\hat{\mu}_4\hat{\mu}_2 + 24\hat{\mu}_2^3}{2\hat{\mu}_6 - 6\hat{\mu}_4\hat{\mu}_2 + 4\hat{\mu}_2^3}
\]
Although the system given by (\ref{eq_4}) has several solutions, it can be shown that (\ref{eq_6}) and (\ref{eq_7}) represent the only real and positive solution for $\hat{K}$ and $\hat{\Gamma}$. 


The performance of proposed estimators $\hat{K}$ and $\hat{\Gamma}$ is investigated by resorting the Monte-Carlo simulation, with the results illustrated in Fig.~\ref{Figure_nova} and   Fig.~\ref{Figure_dd}.
Thereby, for any fixed value of $K$ and $\Gamma$, $500$ realizations of the TWDP process with $N = 10^4$ i.i.d. samples are generated and used in order to determine ${\hat{K}_{j}}$ (\ref{eq_6}) and ${\hat{\Gamma}_{j}}$ (\ref{eq_7}) for $j \in [1, 500]$. These values are then used to calculate sample means  ${\hat{K}}_{mean}$ and ${\hat{\Gamma}}_{mean}$ as ${\hat{K}}_{mean}=(1/500) \sum_{j = 1}^{500} \hat{K}_j$ and ${\hat{\Gamma}}_{mean}=(1/500) \sum_{j = 1}^{500} \hat{\Gamma}_j$. 
Accordingly, Fig.~\ref{Figure_nova} shows the boundaries where the estimates of parameter $K$ from each realization of TWDP process are located in respect to the mean estimated value $\hat{K}_{mean}$, while Fig.~\ref{Figure_dd} illustrates the boundaries where each estimate of parameter $\Gamma$ is located in respect to its mean estimated value.

From Fig.~\ref{Figure_nova} can be observed that the estimator of parameter $K$ given by (\ref{eq_6}) provides accurate results, especially in the region of medium and large values of $K$ and $\Gamma$ (e.g. 
$\Gamma \geq 0.3$ and $K \geq 3$ for $N = 10^4$ samples), for which $\hat{K}_{mean}$ is very close to $K$. 

From Fig.~\ref{Figure_dd} can be observed that estimated values of $\Gamma$, $\hat{\Gamma}_j$, are always smaller than one. Consequently, as $\Gamma$ approaches one, 
$\hat{\Gamma}_{mean}$ starts to increasingly deviate from $\Gamma$, causing increasement in the error of its estimation. 
Accordingly, Fig.~\ref{Figure_dd}  indicates that no anomalies ascertained for $\hat{\Delta}$ can be observed within the $\hat{\Gamma}$.  
It also can be observed that for a small values of $K$ (in the vicinity of one), $\hat{\Gamma}$ provides accurate estimation only in the narrow range of $\Gamma$ values close to $0.6$. However, from the practical point of view, the results for relatively small values of $K$ are irrelevant since in that region TWDP and Rayleigh distributions are almost identical. On the other side, increment of $K$ increases the range of $\Gamma$ values where the estimation is accurate (more precisely, for a considered simulation with $N = 10^4$ samples, $\hat{\Gamma}_{mean}$ is remarkably close to $\Gamma$ for $0.2 \leq \Gamma \leq 0.8$ and $K \geq 3$).
From Fig.~\ref{Figure_dd} also can be observed that in the considered range, dispersion of estimated values is quite insignificant, indicating that derived estimator provides accurate estimates of $\Gamma$ even for relatively small number of samples (i.e. $10^4$).

\section{AsV and CRB}
\label{Sec:3A}
\begin{figure*}[h!]
\centering
\begin{minipage}[t]{1\textwidth}
  \centering
  \includegraphics[trim={0.5cm 0.1cm 1.2cm 1.6cm},clip,width=0.6\linewidth]{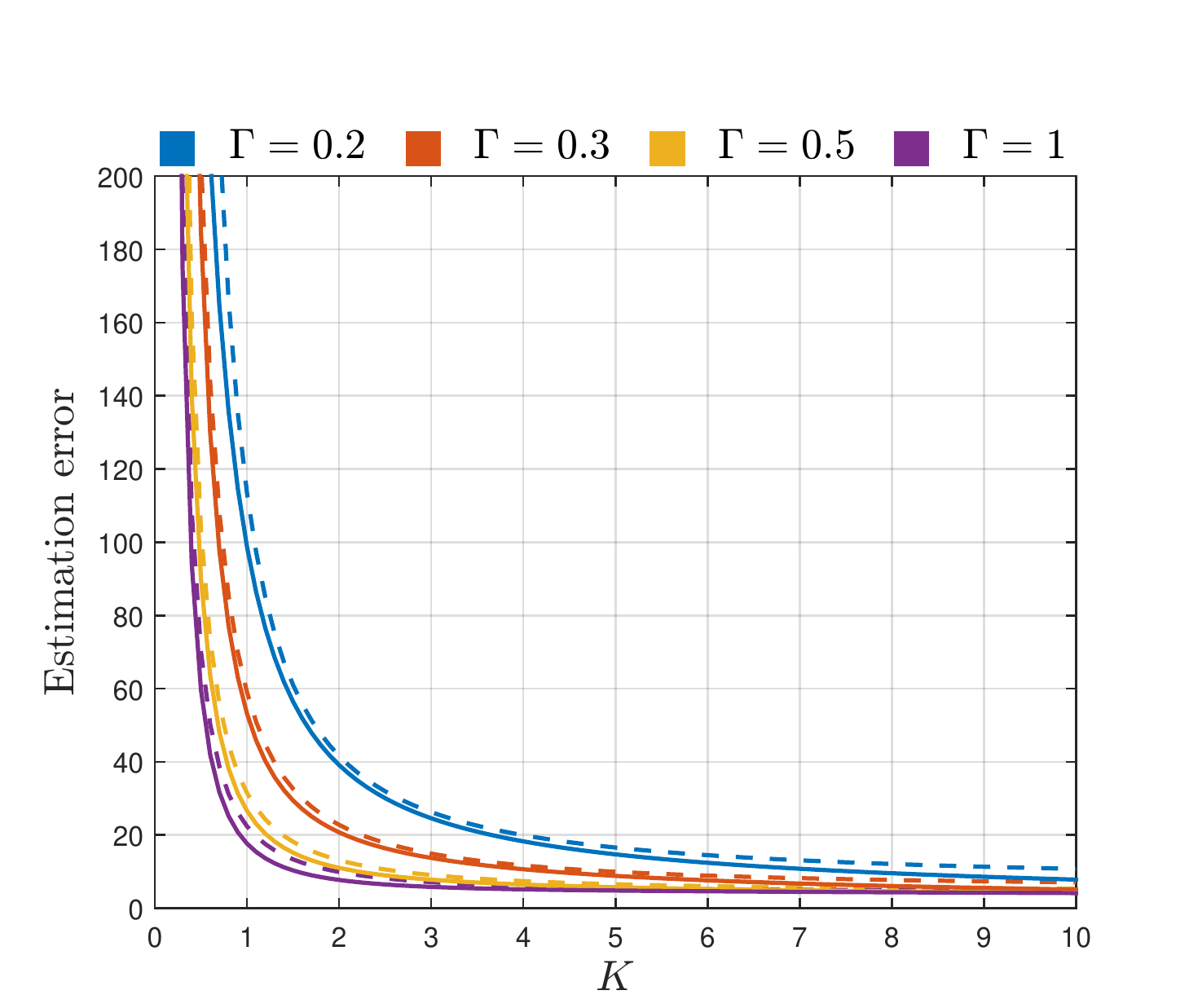}
  \caption{$\sqrt{CRB_K N/K^2}$ (solid line) and $\sqrt{AsV_K N/K^2}$ (dashed line) of $\hat{K}$ given by (\ref{eq_6}), for different values of $\Gamma$}
  \label{Figure1}
\end{minipage}%
\vspace{0.7cm}
\begin{minipage}[t]{1\textwidth}
  \centering
  \includegraphics[trim={0.5cm 0.1cm 1.2cm 1.6cm},clip,width=0.6\linewidth]{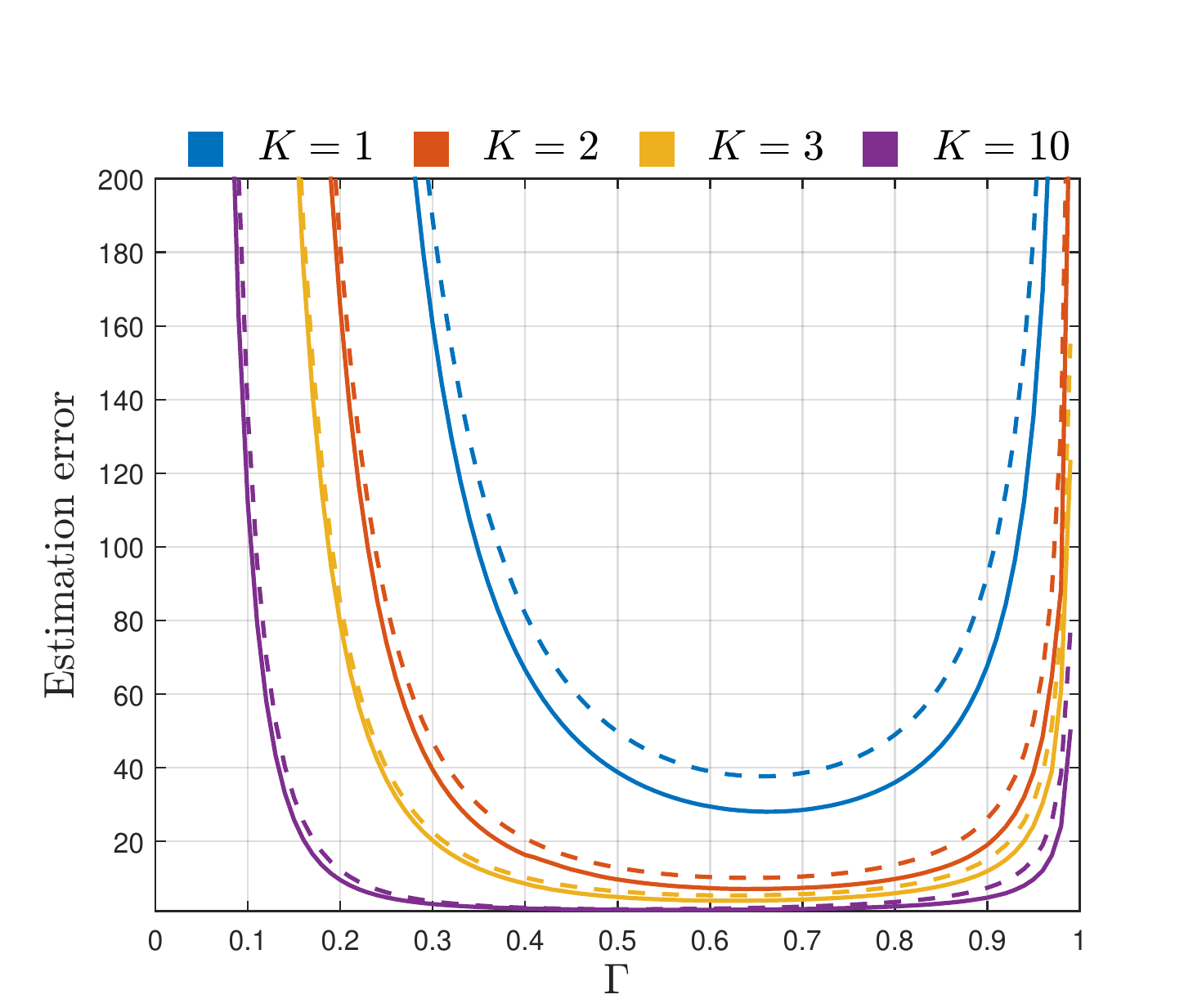}
  \caption{$\sqrt{CRB_\Gamma N/{\Gamma}^2}$ (solid line) and $\sqrt{AsV_\Gamma N/{\Gamma}^2}$ (dashed line) of $\hat{\Gamma}$ given by (\ref{eq_7}), for different values of $K$}
  \textbf{\label{Figure2}}
\end{minipage}%
\end{figure*}

To further assess the performance of the proposed estimators, equations~\cite[eq. (17) - (19)]{Fer16-1} are used to derive corresponding asymptotic variances for $K$ and $\Gamma$, $ASV_K$ and $AsV_{\Gamma}$. Also, the Cramer-Rao lower bounds, denoted as $CRB_{\Gamma}$ and $AsV_{\Gamma}$, are calculated numerically using~\cite[eq. (20)]{Fer16-1}, by employing the Fisher Information Matrix applied on TWDP envelope PDF~\cite[eq. (7)]{rad}, obtained by modifying infinite-series expression originally derived in~\cite{Kos78}. Thereby, following the approach presented in~\cite{Fer16-1, Fer16}, instead of using CRB and AsV, square root CRB and AsV normalized to $N$ and the true value of estimating parameter are used in estimators performance assessment. Thus, sqrt-normalized CRB and AsV of $\hat{K}$, $\sqrt{{CRB_K N}/{K^2}}$ and $\sqrt{{AsV_K N}/{K^2}}$, are plotted in Fig.~\ref{Figure1}, while sqrt-normalized CRB and AsV of $\hat{\Gamma}$, $\sqrt{{CRB_{\Gamma} N}/{\Gamma^2}}$ and $\sqrt{{AsV_{\Gamma} N}/{\Gamma^2}}$, are plotted in Fig.~\ref{Figure2} and denoted as estimation error.

\begin{figure*}[t]
\centering
\begin{minipage}[h]{1\textwidth}
  \centering
  \includegraphics[trim={0.5cm 0.1cm 1.2cm 1.6cm},clip,width=0.6\linewidth]{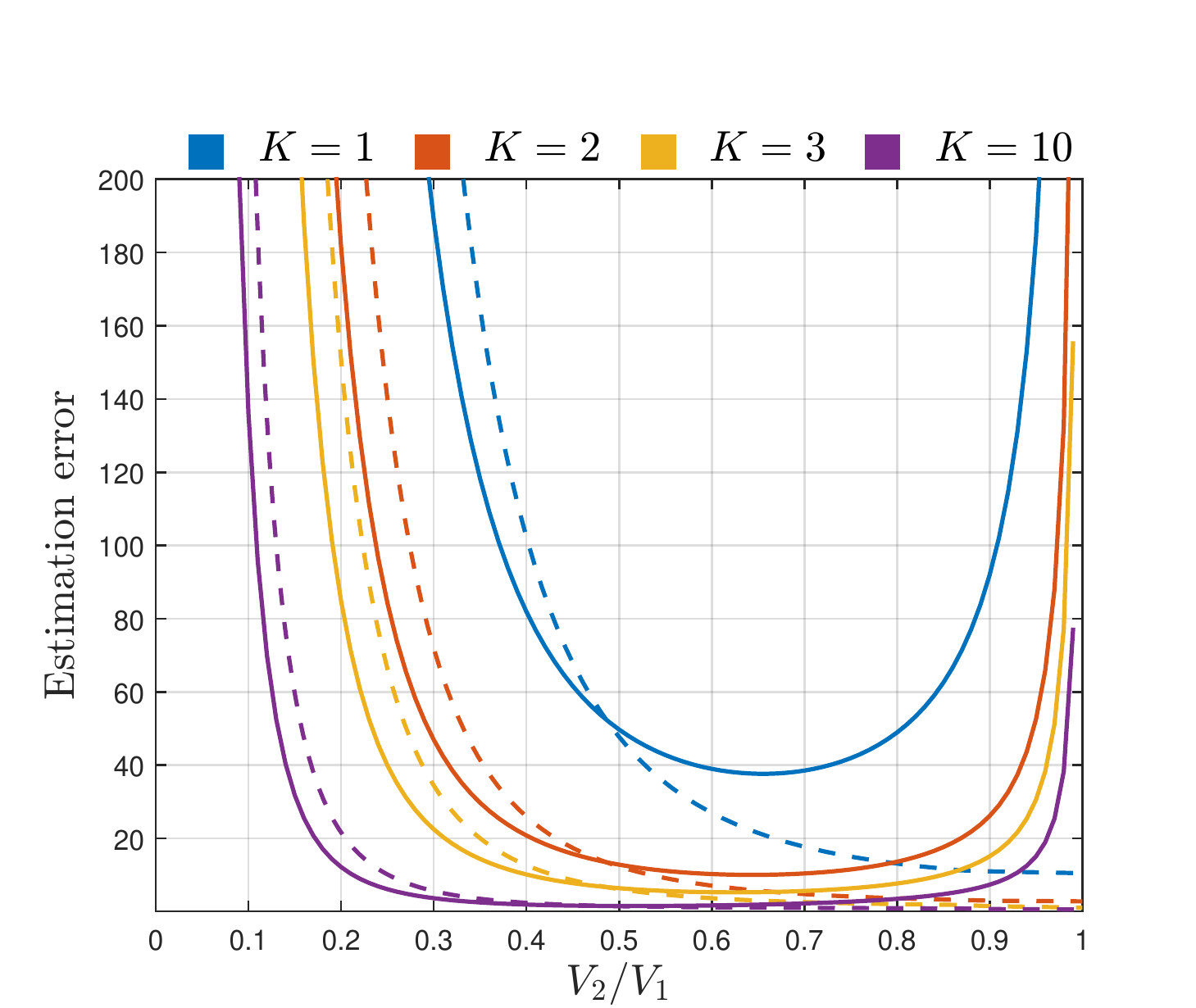}
  \caption{$\sqrt{AsV_\Gamma N/(V_2/V_1)^2}$ of $\hat{\Gamma}$ (solid line) and $\sqrt{AsV_\Delta N/(V_2/V_1)^2}$ of $\hat{\Delta}$ (dashed line), for different values of $K$}
  \label{Figure4}
\end{minipage}%
\end{figure*}

From Fig.~\ref{Figure1} can be observed that the estimation error of $K$ increases with the decreasement of parameter $K$. Thereby, when the  power of specular components $V_1^2 + V_2^2$ is small in respect to the power of diffuse components $2\sigma^2$ (e.g. when TWDP channel becomes Rayleigh-like), error in estimation of $K$ is very large. However, as value of parameter $K$ increases, i.e. as $V_1^2 + V_2^2$ overrides $2\sigma^2$, the estimation of parameter $K$ becomes very accurate. Fig.~\ref{Figure1} also shows that the error in estimation of $K$ grows with the reduction of $\Gamma$, indicating that it becomes harder to accurately estimate $K$ as the specular component with the magnitude $V_2$ becomes more insignificant in respect to one with the magnitude $V_1$.

From Fig.~\ref{Figure1} can also be observed that the values of sqrt-normalized $AsV_K$ are remarkably close to the sqrt-normalized $CRB_K$ for the entire considered range of parameters $K$ and $\Gamma$, indicating almost asymptotic efficiency of  proposed estimator of a parameter $K$. 

Fig.~\ref{Figure2} shows that the estimation error of parameter $\Gamma$ behaves similarly as the estimation error of $\hat{K}$, in respect to $K$ and $\Gamma$. 
Hence, the estimation of $\Gamma$ deteriorates with the reduction of $K$, i.e. as the power of diffuse components becomes more significant in respect to $V_1^2 + V_2^2$. The estimation error of $\hat{\Gamma}$ is large for small values of $\Gamma$, indicating that it is hard to estimate the values of $\Gamma$ when $V_2$ is insignificant in respect to $V_1$. 
For moderate values of $\Gamma$, $\hat{\Gamma}$ given by (\ref{eq_7}) starts to provide pretty accurate results, especially for large values of $K$. However, as $\Gamma$ approaches one, 
estimation of $\Gamma$ becomes more inaccurate. In these conditions,  
the magnitudes of specular components, $V_1$ and $V_2$, become similar. Thereat, since their phase difference is uniform, 
the probability of destructive superposition of specular components becomes pretty large, making their overall magnitude often insignificant. Thus, as $\Gamma$ approaches one, it gets harder to accurately determine the value of $V_2/V_1$, especially when the power of diffuse components is large in respect to $V_1^2 + V_2^2$. 



When it comes to the observation of the proximity of proposed $\Gamma$ estimator to its CRB, form Fig.~\ref{Figure2} can be concluded that the values of the sqrt-normalized $AsV_{\Gamma}$ are remarkably close to the sqrt-normalized $CRB_{\Gamma}$ for $K \geq 2$ in the entire range of $\Gamma$, making the proposed estimator asymptotically efficient for the considered values of $K$.

Accordingly, except providing estimations significantly close to the corresponding CRBs, moment-based estimators (\ref{eq_6}) and (\ref{eq_7}) provide accurate estimates obtained from relatively small number of samples, which can be clearly observed from Fig.~\ref{Figure1} and Fig.~\ref{Figure2}. Namely,
if we assume that the sufficient accuracy of estimation process is achieved when the relative estimation error (obtained by multiplying estimation errors $\sqrt{AsV_K N/K^2}$ and $\sqrt{AsV_{\Gamma} N/{\Gamma}^2}$ by $1/\sqrt{N}$) is smaller than $20\%$, it can be reached by employing only $N = 10^4$ samples for $\Gamma \in [0.3, 0.9]$ and $K \geq 3$.
If necessary, the estimation accuracy in the determined region can be increased, or the region itself can be further expanded, by involving more samples within the estimation process (e.g. by employing $N = 10^6$ samples, relative estimation error in the considered region of a tuple $(K, \Gamma)$ could be reduced to $2\%$, or the estimation error of $20\%$ could be achieved for the wider range of $K$ and $\Gamma$, i.e. $K\geq3$ and $\Gamma \in [0.16, 0.99]$). In this way, the procedures used to create Fig.~\ref{Figure1} and Fig.~\ref{Figure2} can be used to determine the number of samples needed to obtain desired estimation accuracy within the desired range of parameters $K$ and $\Gamma$. 


\vspace{-0.1cm}
\section{Comparison of  conventional and improved moment-based TWDP parameters estimators}
\label{Sec:4}

In order to observe qualitative differences between $\Gamma$- and $\Delta$-based parameterization and to gain more precise insight into the relation between estimation errors for considered parameterizations, $AsV_{\Delta}$ and $AsV_{\Gamma}$ are normalized to the same parameter $V_2/V_1$ and presented in Fig.
\ref{Figure4}. That enables us to compare absolute values of AsV for  $\Delta$ and $\Gamma$ and to discover the differences in their estimation errors for considered ratios between $V_1$ and $V_2$.   
Fig.~\ref{Figure4} shows that for $0\leq V_2/V_1 \leq 0.5$, error in estimation of  $\Gamma$ is two time smaller than the error obtained by estimating $\Delta$. 
On the other side, for $0.5 < V_2/V_1 < 0.8$ and $K \geq 3$, there is no significant difference in accuracy of $\hat{\Delta}$ and $\hat{\Gamma}$.
Finally, for $V_2/V_1 \in [0.8, 1]$, error in estimation of $\Gamma$ starts to increase with the increment of $V_2/V_1$, thus being in line with the model's physical mechanisms. On the contrary, $\Delta$-based parameterization provides fake accurate results, obtained by considering also values of $\hat{\Delta}$ greater than one in calculation of $AsV_{\Delta}$.

Except benefits of $\Gamma$-based parameterization observed in respect to  $\hat{\Gamma}$, it also enables reduction of the estimation error of a parameter $K$, for a much wider set of values of a parameter which reflects the relation between $V_1$ and $V_2$.
Namely, based on the expression of parameter $K$ given in terms of $\Delta$ and the results presented in~\cite[Fig. 2]{rad}, it can be concluded that $K/K_{Rice}$ is almost constant for the entire range of small and medium values of $\Delta$, implying that values of $V_2 \in [0,V_1/2]$ make almost no impact on the value of parameter $K$. That  causes quite pronounced errors in estimation of $K$ for the entire range of small and medium values of $\Delta$ (i.e. $0 \leq \Delta < 0.5$), which can be clearly observed from~\cite[Fig. 1]{Fer16-1}. On contrary, when $K$ is expressed in terms of $\Gamma$, no such anomaly can be observed, as shown in Fig.~\ref{Figure1}. In these circumstances, errors in estimation of $K$ are huge only for small values of $\Gamma$ (i.e. $\Gamma < 0.2$).

\section{Conclusion}
\label{Sec:5}
In this paper, the problem of TWDP parameters' estimation has been investigated in depth. The investigation reveled that the existing moment-based estimators of conventional TWDP parameters $K$ and $\Delta$ are not able to provide accurate estimations for various combinations of their values, due to nonphysical definition of parameter $\Delta$. Accordingly, in this paper, a moment-based estimators for improved, physically justified parameters $K$ and $\Gamma$ are derived. It is shown that derived estimators will provide estimates from $10^4$ samples with the estimation error smaller than $20\%$, when parameters $K$ and $\Gamma$ are in the range $K \geq 3$ and $0.3 \leq \Gamma \leq 0.9$. That indicates that the tuple $(K, \Gamma)$ can be efficiently estimated using derived expressions within the range of these parameters expected to be obtained in mmWave band, even from a relatively small number of samples.  
Since estimators of improved parameters enable us to gain precise insight into the ratios between the two specular and specular to diffuse components of TWDP model in the wide varieties of propagation conditions, simultaneously decreasing their estimation errors in respect to conventional parameterization, it is recommended to adopt parameters $K$ and $\Gamma$ as the only relevant parameters for a description of TWDP fading and to revise the existing measurement-based results related to estimation of TWDP parameters in specific propagation conditions.

\section*{Acknowledgment}
The authors would like to thank Prof. Ivo Kostić for many valuable discussions and advice.



\bibliographystyle{IEEEtran}
\bibliography{Literatura}
\end{document}